\documentclass{article}

\usepackage{arxiv}

\usepackage[utf8]{inputenc} 
\usepackage[T1]{fontenc}    
\usepackage{hyperref}       
\usepackage{url}            
\usepackage{booktabs}       
\usepackage{amsfonts}       
\usepackage{amsthm,amsmath}
\usepackage{mathabx}
\usepackage{algorithm,algorithmic}
\usepackage{nicefrac}       
\usepackage{microtype}      
\usepackage{lipsum}
\usepackage{graphicx}
\graphicspath{ {./images/} }

\title{GRANDPA: GeneRAtive Network sampling using Degree and Property Augmentation applied to the analysis of partially confidential healthcare networks}

\author{
 Carly A. Bobak \\
  The Dartmouth Institute for Health Policy and Clinical Practice \&\\ The Department of Biomedical Data Science \& \\ Research Computing\\
  Dartmouth College\\
  Hanover, NH 03755 \\
  \texttt{Carly.A.Bobak@dartmouth.edu} \\
   \And
 Yifan Zhao \\
   The Dartmouth Institute for Health Policy and Clinical Practice\& \\ The Department of Biomedical Data Science\\
 Dartmouth College\\
  Hanover, NH 03755 \\
  \texttt{Yifan.Zhao.GR@dartmouth.edu} \\
  \And
 Joshua J. Levy \\
  Department of Pathology and Laboratory Medicine \& \\ Department of Dermatology \& \\ Department of Epidemiology\\
 Dartmouth College\\
  Hanover, NH 03755 \\
  \texttt{Joshua.J.Levy@dartmouth.edu} \\
    \And
 A. James O'Malley\\
   The Dartmouth Institute for Health Policy and Clinical Practice \& \\ The Department of Biomedical Data Science\\
 Dartmouth College\\
  Hanover, NH 03755 \\
  \texttt{James.OMalley@dartmouth.edu} \\
}

\begin{document}
\maketitle
\begin{abstract}
Protecting medical privacy can create obstacles in the analysis and distribution of healthcare graphs and statistical inferences accompanying them. We pose a graph simulation model which generates networks using degree and property augmentation (GRANDPA) and provide a flexible R package that allows users to create graphs that preserve vertex attribute relationships and approximating retaining topological properties observed in the original graph (e.g., community structure). We support our proposed algorithm using a case study based on Zachary's karate network and a patient-sharing graph generated from Medicare claims data in 2019. In both cases, we find that community structure is preserved, and normalized root mean square error between cumulative distributions of the degrees is low (0.0508 and 0.0514 respectively).
\end{abstract}


\section{Introduction}

Graph theory and network analysis play vital roles in the study of complex relationships relevant to biology, disease, and healthcare delivery. Common structures of these networks include network vertices as genes, proteins, physicians, hospitals, etc. where edges  often represent co-occurrence or correlation \cite{Hu2021,Fernandez-Pena2022TheProtocol,Koutrouli2020,VanDerWijst2018,Infante2020NetworkDisease}. Such networks frequently contain multiple characteristics on the vertex level (referred to as attributes or properties) which contain information on the individual actors, whereas characteristics on the edge level contain information between two actors. For instance, network analysis has been used to study the diffusion of COVID-19 infection through hospital employees \cite{Garzaro2020COVID-19Italy}, to uncover new findings in the pathogenesis of Tuberculosis \cite{Bobak2022IncreasedTuberculosis}, and are frequently used to study and propose novel cancer biomarkers \cite{Kosvyra2021NetworkReview}.

Networks for healthcare applications are often constructed using confidential medical data. Social healthcare networks can be constructed from insurance claims, medical records, and electronic health records \cite{Landon2018Patient-SharingBeneficiaries,Cusumano-Towner2013AData}. Biological networks have traditionally been constructed by observing relationships between potential biomarkers (genes, proteins, transcription factors, etc.) across an entire cohort, but in the era of personalized medicine, networks constructed from one individual's personal biological information have been proposed  \cite{Koutrouli2020,VanDerWijst2018,Infante2020NetworkDisease}. In both social and biological networks, necessary privacy and confidentiality precautions, including storage of network data on secure servers and limitations to thoroughly vetted computational tools of such data, can create obstacles in the analysis and distribution of healthcare networks \cite{Clayton2019TheLimitations,Sathanur2017}. Thus, there is a need to generate novel networks which maintain the structural properties of healthcare networks without compromising or distributing confidential medical data. 

Among the many models and approaches to generate networks that have been developed \cite{Barabasi1999EmergenceNetworks,Erdos1984OnGraphs,Watts1998CollectiveNetworks,Hunter2008Ergm:Networks,Csardi2014,Chandrasekhar2018ASubgraphs}, most emphasize simulating the overall network topology and rarely consider the role of vertex attributes. Exponential random graph models (ERGMs) are among the more flexible options \cite{Hunter2008Ergm:Networks}, although have been noted to have unstable parameter estimation on large networks and those with dyadic depedent terms \cite{Chandrasekhar2018ASubgraphs}. In many cases, the information stored in the vertex attributes is directly related to the application of interest – for instance, 1) studies of homophily (i.e., do vertices with similar attributes connect more frequently than expected after conditioning on other network features and properties), 2) heterophily (i.e., do differing vertices connect more frequently than expected), and 3) studies examining network characteristics related to community or clustering, the spread of health technologies, etc. All of these studies would require the preservation of vertex-level information. To address this, Kim \& Lekovec introduced the Multiplicative Attribute Graph (MAG), which assumes that vertex attributes are indicative of latent graph structure, and seeks to generate graphs using such structure \cite{Kim2010}. Building on this, Pfeiffer et al. introduced the Attributed Graph Model (ATG) which similarly seeks to generate graphs from attribute structure, but uses an accept-reject sampling procedure to do so \cite{Pfeiffer2014AttributedAttributes}. It has been previously posited that both vertex similarity (homophily) and vertex popularity (degree) should be used to generate networks reflective of those observed in the real world \cite{Papadopoulos2012}. In response, Sathanur et. al introduced the Property Graph Model (PGM) which calculates the joint label assignment probabilities for vertices and joint distribution probabilities over pairs of vertices for the edges. They then 'augment' their original label categories by partitioning the degree distribution into bins and assigning each vertex a label corresponding to the bin its degree is in. They demonstrate that this algorithm is scalable (linear in the number of edges), preserves attribute relationships, and better represents degree structure \cite{Sathanur2017}. 

Community detection in networks has long been established as an important component of network study \cite{Newman2006ModularityNetworks,Cherifi2019OnOpportunities}. In real-world networks, nodes naturally organize into clustered communities or modules and these communities are often meaningful units for analysis. For instance, community structure is often considered in biological networks and facilitates the study of how groups of biomarkers work in tandem to control biological processes \cite{Sah2014ExploringGraphs,Alcala-Corona2016CommunityNetwork,Langfelder2008WGCNA:Analysis}. As well, community structure of physicians at hospitals has been associated with patient readmission rates and overall hospitalization costs \cite{Uddin2015ImpactOutcomes}. Hence, graph generating algorithms which preserve community structure alongside vertex attributes are necessary to simulate networks which can be analyzed to study healthcare phenomena. 

In this work, we aim to create an R package to allow for the easy implementation of the PGM generative graphs in the R programming language, favored by many social network scientists. Additionally, we build upon the framework proposed by Sathanur et. al by augmenting the algorithm with an optional community label and creating a flexible framework that allows researchers to embed supplemental information about network structure (such as centrality measures, network geometry, etc.) into graph generation. Such generated networks can be distributed and analyzed without confidentiality concerns as they no longer contain potentially sensitive and identifiable real-world patient or physician data.

\section*{Methods}

\subsection*{Property Graph Models with Structural Augmentation}

Full details on the PGM procedure can be found in Sathanur's original manuscript \cite{Sathanur2017}. In short, let $$ G_s = <V_s,E_s,L, L(V_s)> $$ be a property graph, where $V_s$ is the set of all vertices ($v_i$) and $E_s \subseteq V_s \times V_s$ is the set of all edges. $L=\{L_k\}^M_{k=1}$ is a set of M vertex label sets; and $L_k$ is the set of all possible values for the $k^{th}$ label such that $n_k=|L_k|$. Then $L(V_s)$ is the set of all label value vectors in 1-to-1 correspondence to $V_s$. Thus, each $\bar L(v_i)$ is drawn from the set of all possible joint label assignments $\mathcal{L}=\bigtimes_{k=1}^ML_k$ based on our label vector. There are $N=\prod_{k=1}^M n_k$ possible joint label categories. The $j^{th}$ joint label category is denoted $c_j$. The probability of drawing a joint label category $c_j$ is denoted $P_L(c_j)$ and defined as follows using the observed vertex labels in G:

$$P_L(c_j)=\frac{\sum_{i=1}^{|V_s|}1_{c_j}(\bar L(v_i))}{|V_s|}$$
which uses the following indicator function:
$$1_{c_j}(\bar L(v_i))=\begin{cases}
1,& \text{if   } \bar L(v_i)=c_j\\
0,& \text{otherwise}
\end{cases}$$

Edge connectivity is modeled using a joint distribution over pairs of label categories $(c_j,c_{j'})$; which we denote $P_c$. This probability is calculated as:
$$P_C(c_j,c_j')=\frac{\sum_{<v_i,v_i'\in E_s>}1_{c_j,c_j'}(\bar L(v_i),\bar L(v_i'))}{|E_s|}$$
using  the following indicator function:
$$1_{c_j,c_j'}(\bar L(v_i),\bar L(v_i'))=\begin{cases}
1,& \text{if   } \{v_i,v_i'\}\in E_s \text{ and    } \{\bar L(v_i),\bar L(v_i')\}=\{c_j,c_j'\}\\
0,& \text{otherwise}
\end{cases}$$

We then leverage $P_L$ and $P_C$ to generate a target graph, $G_T$ following Algorithm 1.

\begin{algorithm}[h]
\caption{PGM Algorithm\label{PGM}}
\begin{algorithmic}[1]
\STATE Calculate $P_L$ and $P_C$ from the observed data. \\
\STATE Sample $n_t$ (the number of vertices in the generated graph) from $P_L$. \\
\STATE Add all $n_t$ vertices to an empty network\\
\STATE Create a map, \emph{C2V} which contains a list of all sampled vertices for each category combination (of $N$ combos) \\
\STATE Sample $m_t$ edges from $P_C$.  \\
\STATE For each edge draw an appropriate $v_i$ and $v_i'$ from \emph{C2V} and form an edge.\\
\STATE Return $G_T$
\end{algorithmic}
\end{algorithm}

This procedure guarantees the preservation of $P_L$ and $P_C$ between the original and generated networks, provided the generated network is constructed with the same number of vertices and edges. In the case where the desired generated network has a different size from the original, $P_L$ and $P_C$ are preserved approximately due to rounding error. 

To augment with degree, Sathanur et. al. proposed calculating vertex degree $\forall v_i \in G_s$, and assigning a new label by dividing the degree distribution into $n_b$ bins, where $n_b$ is tuned according to some error metric (for instance, the normalized root mean square error between the distribution of network statistics, motif-based measurements, etc.) between graphs. The degree augmentation algorithm is outlined in Algorithm 2. 
\begin{algorithm}[!h]
\caption{Degree Augmentation Algorithm\label{degAug}}
\begin{algorithmic}[1]
\STATE Check if error $>$ tolerance \\
\STATE Add 1 to $n_b$ \\
\STATE Divide degree range in $G_s$ into $n_b$ intervals and assign to new attribute $l_a$ \\
\STATE Append label vector $\bar L(v)$ with $l_a(v)$ \\
\STATE Simulate new graph using PGM algorithm \\
\STATE Compute new error. Repeat while error $>$ tolerance. \\
\end{algorithmic}
\end{algorithm}

It has previously been shown that community detection as well as degree is beneficial in graph generation \cite{Karrer2010}. Thus, we sought to extend the above framework to consider a community detection augmentation option. To achieve this, we added additional labels $l_c$ to represent categorical, non-overlapping communities and appended these to the attribute labels $L(v)$. These labels are then processed alongside the vertex and edge probabilities for attributes in $P_L$ and $P_C$. This framework is flexible; and hence allows users to add additional structural information as label categories by creating additional labels $l_{struct}$, adding these to $L_v$ and calculating subsequent node and edge probabilities. We call our method GeneRAtive Networks with Degree and Property Augmentation (GRANDPA). Algorithmically, we propose the following changes to the PGM fitting procedure:

\begin{algorithm}[h!]
\caption{GRANDPA Algorithm\label{gramps}}
\begin{algorithmic}[1]
\STATE Fit community detection model on network and tune appropriately \\
\STATE Add an attribute label based on community membership $l_c$\\
\STATE Append label vector $\bar L(v)$ with $l_c(v)$ \\
\STATE Simulate new graph using PGM algorithm (Algorithm \ref{PGM}) \\
\STATE Compute new error. Check error $>$ tolerance. \\
\STATE Run degree augmentation algorithm (Algorithm \ref{degAug})
\end{algorithmic}
\end{algorithm}

\begin{figure}[h!]
    \centering
    \includegraphics[width=0.95\linewidth]{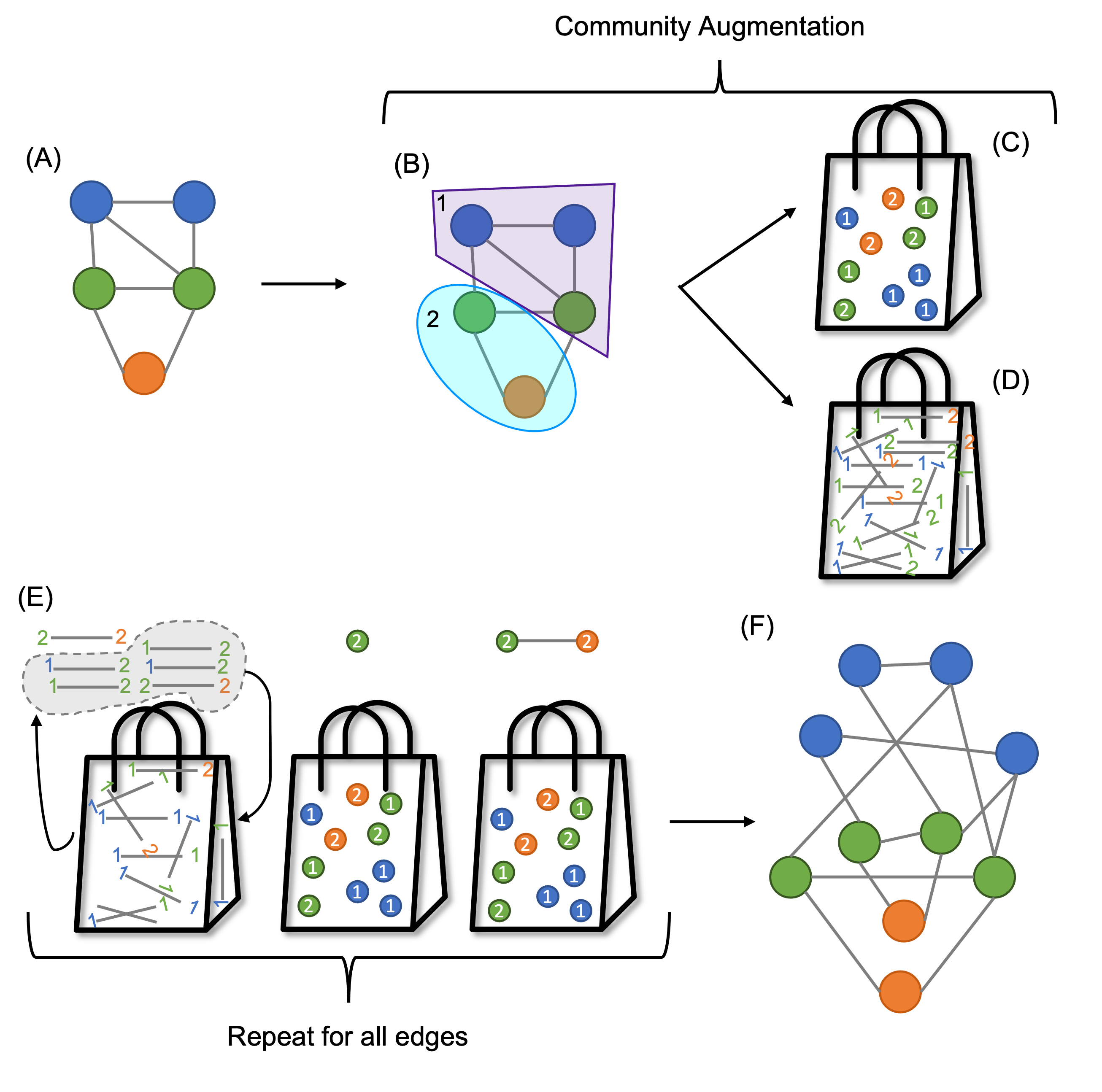}
    \caption{A depiction of the GRANDPA methodology used to generate graphs. (A) Shows an example original network, (B) shows a community calculation on this network, (C) and (D) show the generation of bags of possible vertices and nodes using $P_L(c_j)$  and $P_c$ respectively. (E) Demonstrates the sampling procedure, wherein edges are sampled without replacement. (F) is an example generated graph.}
    \label{fig:Method}
\end{figure}

The entire GRANDPA procedure is depicted in Figure \ref{fig:Method}.

\subsection*{GRANDPA: A User-friendly Implementation in R}

GRANDPA is available as an R package located at \url{https://github.com/CarlyBobak/grandpa}. GRANDPA depends exclusively on the \textit{igraph} framework \cite{Csardi2014} and calculates probabilities using \textit{tidyr} and \textit{dplyr} \cite{Wickham2019WelcomeTidyverse}. To use GRANDPA, users feed the function an \textit{igraph} object, where any attributes which should be used to generate the joint label distribution should contain the word "label". Thus, any vertex metric could be flexibly added to the label space to be used as part of the network generation. The source code was developed and analyses conducted in R version 4.1.1 \cite{r}.

\subsection*{Case Study: Zachary Karate Network}

Contained within the \textit{igraph} package is a small example network from Zachary's 1977 paper studying conflict in a Karate network \cite{Csardi2014,Zachary1977AnJSTOR}. We created two node category labels associated with network structure by using sampling weights to fix features whose distribution we wanted to retain while allowing for natural variation in the status of any given actor and edge. The first label has three levels and its sampling weight is correlated with the degree distribution of the graph. For the second attribute, four categories are assigned depending on the distribution of closeness centrality. To easily assign sampling probabilities, we computed the quantiles of the degree distribution and closeness centrality scores as cutoffs, and assigned labels to nodes with varying sample probabilities based on those cutoffs. This sampling procedure ensures that the expected values of the concerned statistics are retained while allowing inherent noise between our assigned labels and the network structure, providing the opportunity to test whether attribute labels contribute to and are recovered by the networks constructed from our generative network modeling process when additional structural information is augmented.

To augment generated graphs with degree distributions, we tuned $n_b$ to be optimized over the values of 3, 5, 7, 10, and 15, and compared the normalized root mean square error (NRMSE) between the complementary cumulative distribution function (CCDF) of the vertex degrees between the original and generated graphs. We also calculated communities in the original graph using the edge betweenness algorithm \cite{Newman2004FindingNetworks}. We used the same algorithm to identify communities in generated graphs, and systematically compared both the number and size of communities between the returned graphs.  

\subsection*{Case Study: Unipartite Medicare Network}

Bipartite networks can be constructed from insurance claims, such as those generated through the Centers for Medicare \& Medicaid Services (CMS) Program \cite{Landon2018Patient-SharingBeneficiaries}. Networks connect patients to each physician they file a claim with over a designated period of time. Such networks can be projected to a unipartite space, where physicians are connected if they shared patients over a period of time \cite{Barnett2011MappingData}. To demonstrate the potential of GRANDPA in social network analysis for healthcare applications, we constructed a unipartite patient sharing network graph using Medicare Data from 2019, where two physicians are connected if they shared at least 11 patients. Such graphs are large, containing many nodes and edges, and are difficult to visualize \cite{Allen2020HairballGraphs}. For demonstration purposes, we subset this graph by randomly selecting 3 physicians, finding all third-degree neighbors, and including those physicians and neighbors for our graph generation. 

Each physician in our data had a known primary medical specialty. We calculated community membership for each physician which may be indicative of hospital affiliation, working groups, geographical regions, etc. using the fast-greedy algorithm in igraph \cite{Csardi2014,Clauset2004FindingNetworks}. Additionally, we calculated the linchpin centrality of each physician based on their connections to communities outside their own \cite{Nemesure2021}, and binned linchpin centrality into labels with no centrality (=0), low centrality (between 0 and 0.2), or high linchpin centrality ($\geq 0.2$). We augmented degree by increasing the number of bins in intervals of 5. 

To compare graphs, we plotted the CCDF and calculated the NRMSE between the original source graph and the generated graphs. We also ran the fast-greedy community detection algorithms on the generated graphs and selected graphs which approximated both the size and number of returned communities. We ran a binomial regression on the original graph which aimed to predict if a physician's primary specialty was hospitalist or internal medicine using a homophily variable (same specialty). We repeated this analysis on our best-generated graph, and compared the returned coefficients (odds ratio) to evaluate if models built on the simulated graph reflect the results of models built on the original. 

\section*{Results}

We use two real-world networks to demonstrate the utility of the GRANDPA algorithm, particularly in graphs with inherent community structure. The first is the canonical Karate network discussed by Zachary \cite{Zachary1977AnJSTOR}. The second is a unipartite patient-sharing network which links healthcare providers who shared Medicare patients in 2019.

\begin{figure}[h!]
    \centering
    \includegraphics[width=0.85\linewidth]{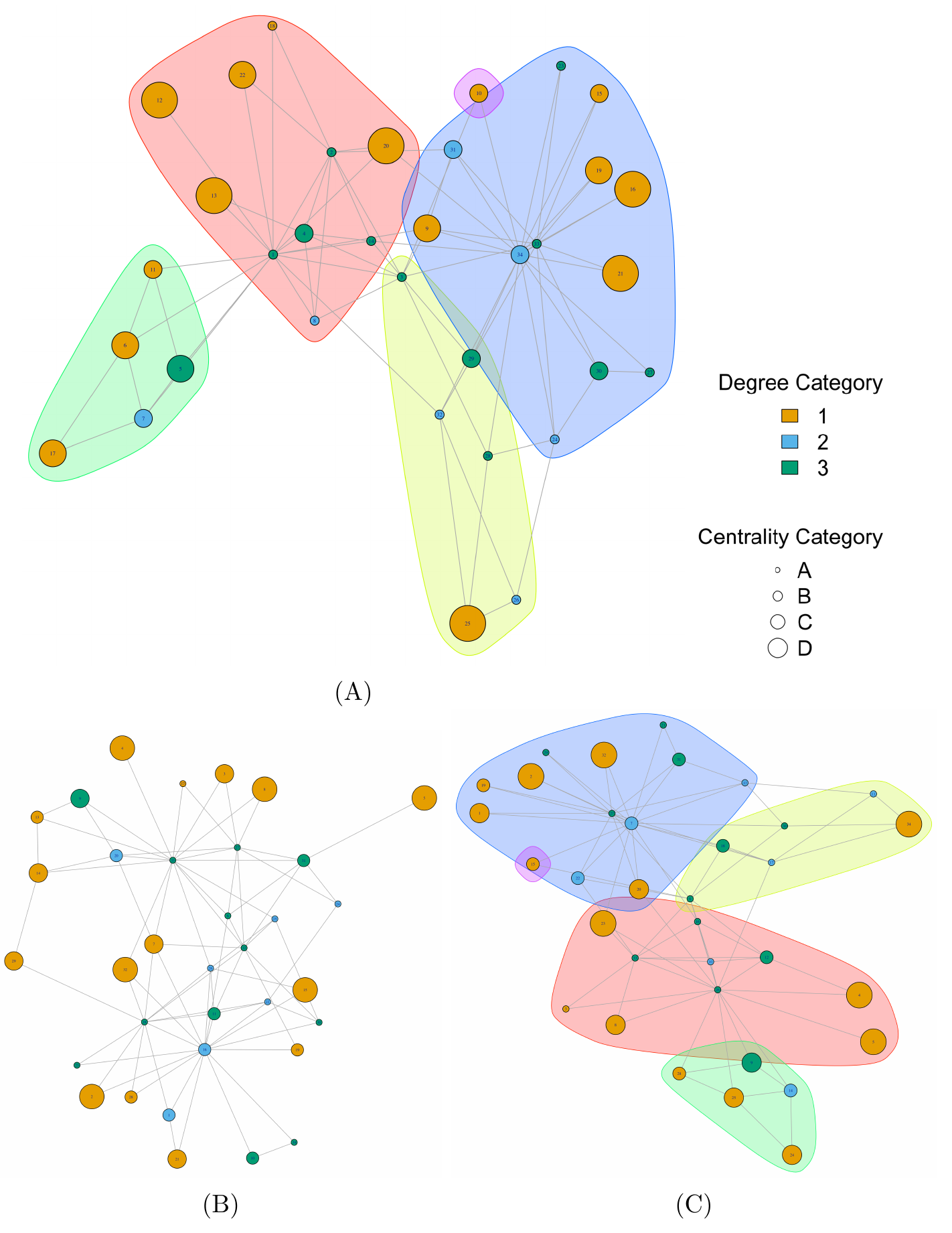}
    \caption{The original and returned graphs from Zachary's Karate network\cite{Zachary1977AnJSTOR}. (A) The original graph, where vertex color indicates a category associated with degree, vertex size represents a category associated with centrality, and community areas are drawn. (B) The returned graph, if just categorical attributes and degree augmentation are used to generate the network. (C) The generated graph, where attributes, community labels, and degree are used to generate the network.}
    \label{fig:Karate}
\end{figure}

\subsection*{Case Study: Zachary Karate Network}\label{lab:CS1}
 
To illustrate the generality of our GRANDPA algorithm, we first chose the network of a university karate club. The karate network consists of 34 vertices with 78 undirected edges. Each node represents a club member, and an edge between two members indicates their interaction. 

The plots of the three selected graphs are shown in Figure \ref{fig:Karate}. To better visualize the correlation between network structure and node attributes, we selected colors to depict the different levels of the first label, and sized nodes to show the corresponding categories of the second one. The original graph, with generated labels and communities, is shown in Figure \ref{fig:Karate}.

\begin{figure}[t!]
    \centering
    \includegraphics[width=0.6\linewidth]{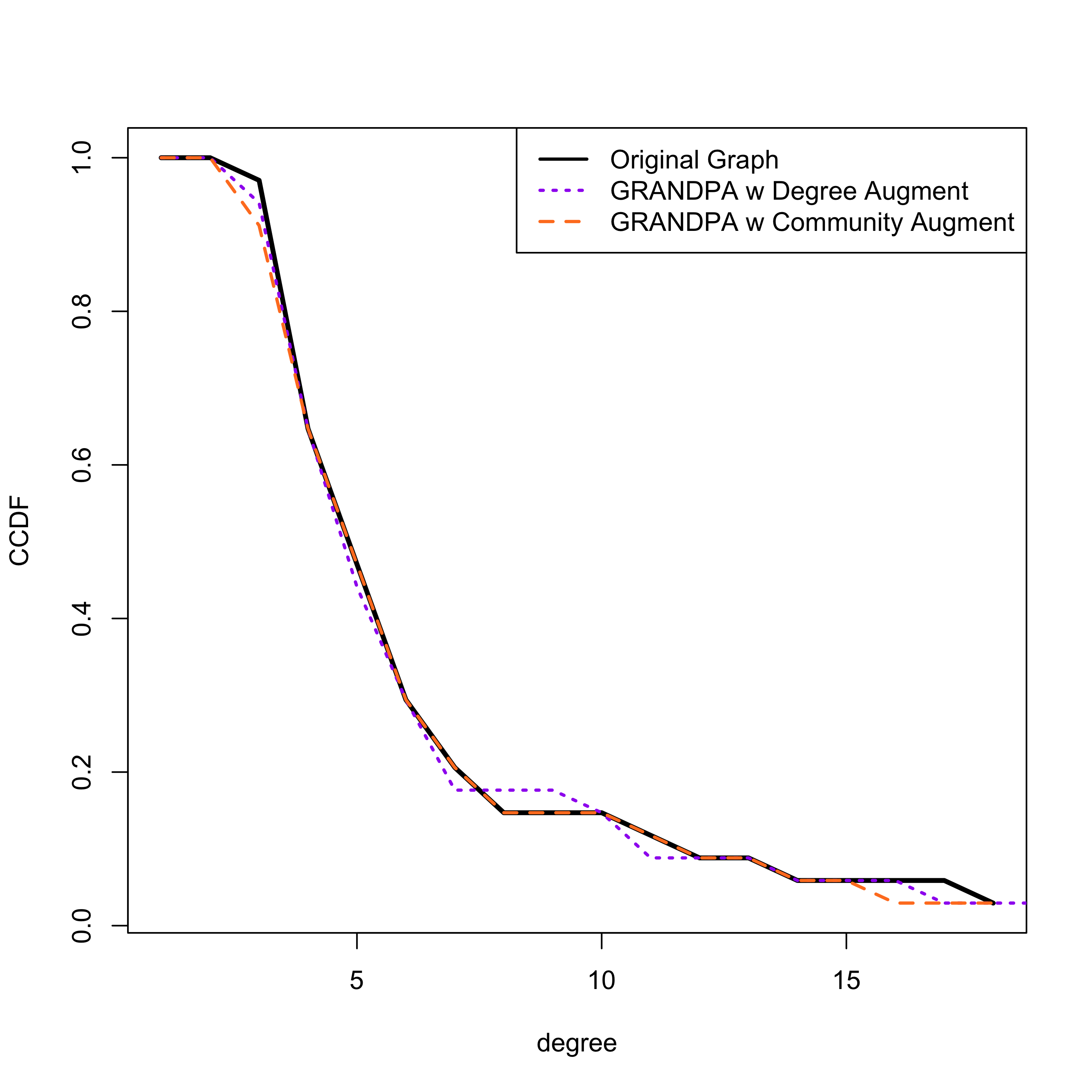}
    \caption{The CCDF of the vertex degree between the original and generated networks for Zachary's Karate data\cite{Zachary1977AnJSTOR}}
    \label{karateAcc}
\end{figure}

Two generated graphs are shown. The first represents a generated graph with both attributes and augmented degree ($n_b$=10). The overall degree distribution across the vertices is close with $NRMSE=0.161$ and shown in Figure \ref{karateAcc}. However, visually it is clear that the community structure present in the original graph is missed in the generated graph. 

When we create a community label following the marked areas and feed this to the GRANDPA algorithm, we quickly recover a graph that is highly similar to the original, which is shown in \ref{fig:Karate}. The NRMSE between the degree distribution of the original and returned graph with community augmentation is nearly identical, as shown in Figure \ref{karateAcc} with overall value $NRMSE=0.0508$. As well, we recaptured nearly identical communities, in both overall number of communities, size of communities, and internal community structure. 

\subsection*{Case Study: Medicare Unipartite Physician Network}\label{lab:CS2}

\begin{figure}[h!]
    \centering
    \includegraphics[width=0.65\linewidth]{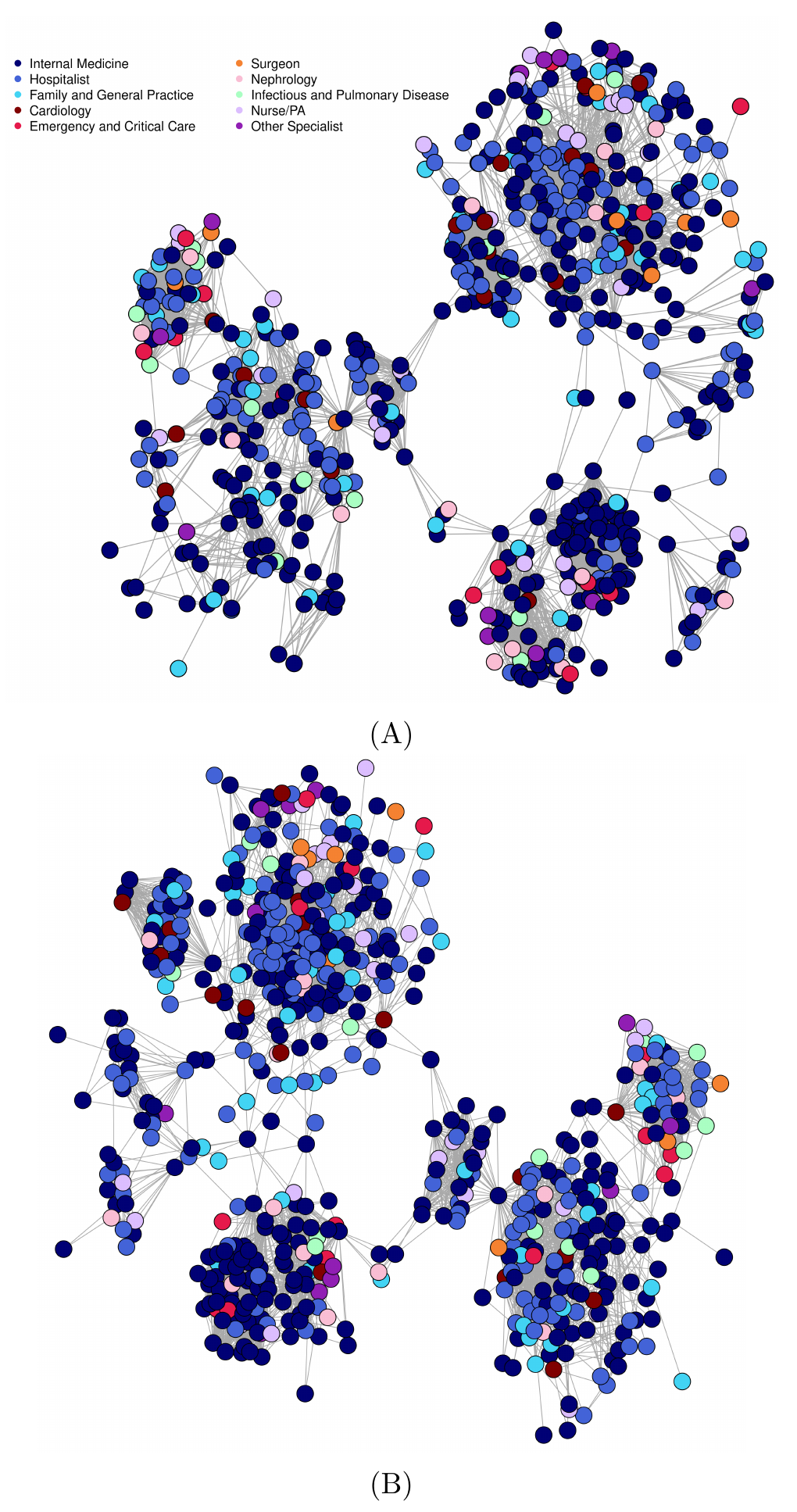}
    \caption{The original and returned graphs from our 2019 Medicare data. (A) The original graph, where vertex color indicates a physician's primary specialty. (B) The final returned graph with community, degree, and linchpin centrality augmentation.}
        \label{fig:CMS}
\end{figure}

We sought to test our algorithm on a large, heterogeneous healthcare network which is more representative of those used in biomedicine and healthcare policy studies. We constructed a network using Medicare claims data from 2019  (8,021 vertices, 82,893 edges) and subset it to include 608 physicians (vertices) and 6,480 patient-sharing connections (edges). The original graph is shown in Figure \ref{fig:CMS}. Vertex colours are representative of the physician's primary specialty. Community-level clustering is present, and reflective of hospitals, geographic regions, and physician working groups. A fast-greedy clustering algorithm detected 10 disparate communities with sizes $\{209,134,105,41,34,32,21,14,11,7\}$.  

\begin{figure}[t!]
    \centering
    \includegraphics[width=0.6\linewidth]{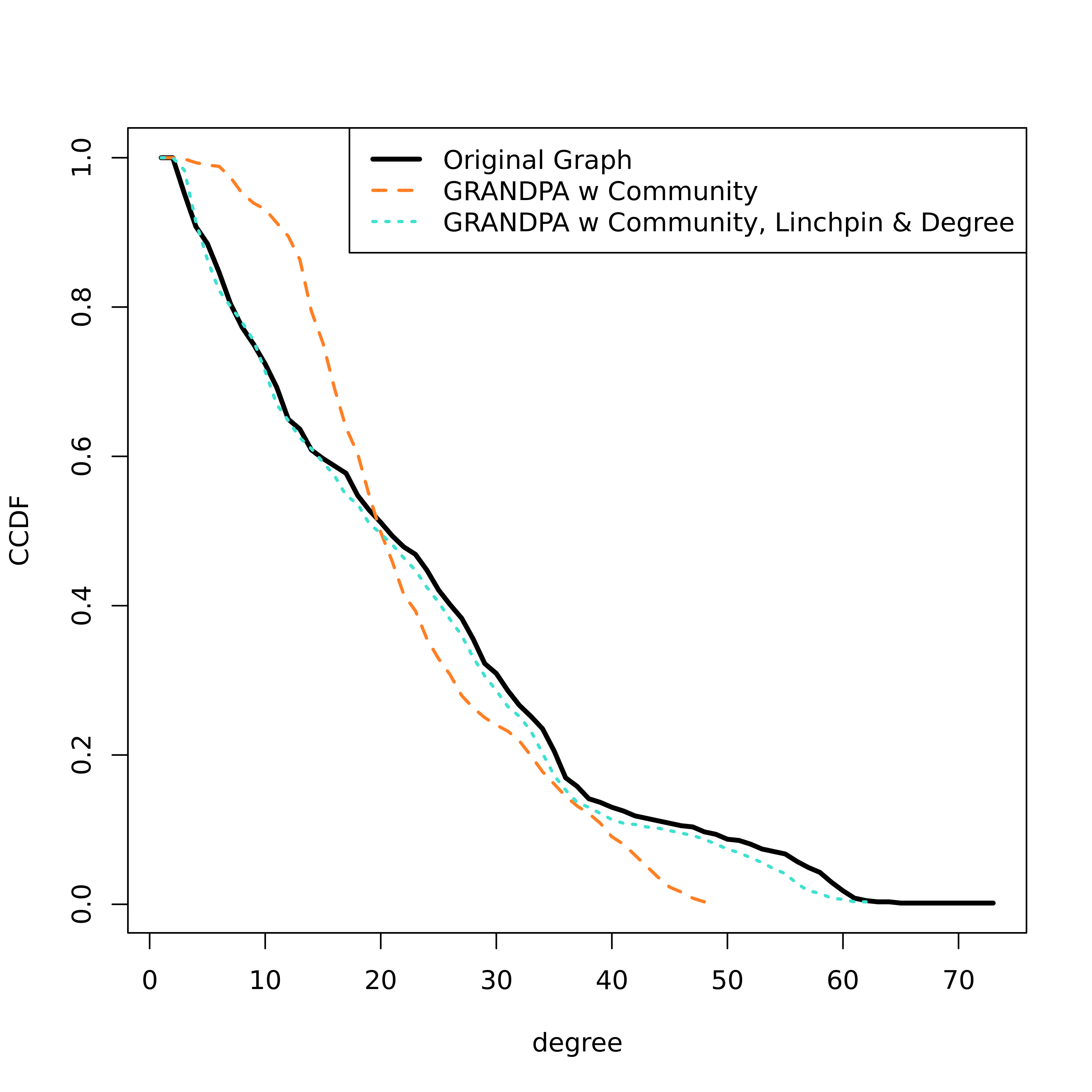}
    \caption{The CCDF of the vertex degree between the original and generated networks for the 2019 Medicare Data}
    \label{CMSAcc}
\end{figure}

Graphs were generated using just attribute-level information (Specialty), attribute information with degree augmentation, attribute information with community augmentation, and attribute information with community augmentation and degree augmentation. Similar to the Karate Case Study, generated graphs without community augmentation lacked the community structure observed in the original graph. While the generated graph with only attribute information and community augmentation produced a graph with matching communities, the degree distribution was poorly recovered (NRMSE=0.237; Figure \ref{CMSAcc}). Augmenting further with a degree label ($n_b=15$) reduced the NRMSE to 0.054. However, the generated graph did not contain key characteristic nodes connecting communities (often referred to as "bridge" nodes \cite{Ezeh2019Multi-typeCommunities}) as was observed in the original graph. 

To attempt to ameliorate the lack of recovery of bridge nodes, we used Linchpin Centrality score \cite{Nemesure2021} to identify nodes which are 'one-of-a-kind' compared to their neighbors' community labels, and further augmented our GRANDPA algorithm with a label corresponding to linchpin centrality. The generated graph with community, degree and linchpin centrality augmentation is shown in Figure \ref{fig:CMS}. The NRMSE between the CCDF of the final graph and the original was 0.051 (Figure \ref{CMSAcc}). Similar to the original graph, the final graph had 10 identifiable communities with sizes $\{205,132,108,42,35,33,21,14,11,7\}$, showing a high degree of concordance with the community structure of the original graph.  

We sought to evaluate if models trained on the generated graph were reflective of results trained on the original graph. To this end, we trained a regression model on both the original graph and the final graph (Figure \ref{fig:CMS}) which aimed to predict if a physician's primary speciality was either hospitalist or internal medicine given the proportion of their neighbors with the same specialty. In the original graph, the odds ratio that a physician is a hospitalist or internal medicine doctor is 2.486 (95\% confidence interval: 2.275, 2.716) when the proportion of physician neighbors within the same specialty changes from 0 to 1. In the final generated network, the same odds ratio is 2.705 (95\% confidence interval: 2.478, 2.952). We generated 10 graphs with different random seeds and recalculated the odds ratio for each graph. The mean odds ratio in the generated graphs was 2.703 with a standard deviation of 0.023. 

\section*{Discussion}

Data sharing and availability are fundamental to reproducibility of healthcare research and trust in the results obtained. However, concerns on protecting patient privacy need to be balanced \cite{Wirth2021Privacy-preservingComparison,Clayton2019TheLimitations,Hammack-Aviran2020ResearchPermission,McGraw2021PrivacySystem}. This is also true for the analysis and distribution of networks generated using patient health data. The American Medical Association recently surveyed patients on the use of their medical data for research, and 75\% of respondents wanted the ability to opt-in for research use. As well, 92\% of respondents felt that their data should not be available for purchase by either corporations or individuals \cite{MedicalAssociation2022PatientPage}. Separately, Hammack-Aviran et al. noted in a qualitative survey that patients would prefer transparency and choice in having their electronic medical records included in research efforts, often wanting information prior to consent on all research objectives and stakeholders \cite{Hammack-Aviran2020ResearchPermission}. Moreover,  open-source software is often desirable for analyzing patient data, but may have security vulnerabilities which violate HIPAA compliance \cite{Farhadi2019ComplianceRequirements}. Assessing open-source software for compliance is a non-trivial task \cite{Farhadi2019ComplianceRequirements}. Researchers are required to de-identify data prior to distribution. However, in the case of biological data, such as data generated from DNA, data itself is a unique identifier \cite{Clayton2019TheLimitations}. Moreover, patients often view their genetic and molecular data to be private \cite{Clayton2019TheLimitations}. Patient trust is a necessary component in healthcare research, and hence the distribution and analysis of networks with embedded patient data needs to be done with extreme care. Simulation of networks can alleviate many of these concerns as it allows for the creation of networks which maintain macro-level relationships without revealing any individual-level observations.

In this work, we demonstrated that GRANDPA can be used to generate both simple and complex graphs which are highly representative of original real-world networks. The GRANDPA framework allows the implementation of a family of methods, in which researchers can customize information pertaining to attributes and network structure to generate realistic graphs. Both case studies reproduced graphs with highly similar community structure and degree distributions while recovering the relational attribute structure by design. In our two case studies, degree distributions were highly preserved, and community detection algorithms identified nearly identical communities between the original and generated graphs.

Our case study of Zachary's karate network reproduced a graph that is highly representational of the original graph, with many motifs within communities preserved between the original and generated graphs. Our case study of a patient-sharing network generated from CMS data likewise demonstrated that community structure could be preserved, and also that regression coefficients generated on the original and generated graphs overlapped, suggesting that models fit on the generated graphs are reflective of real-world findings. 

We demonstrated that the GRANDPA algorithm can flexibly be augmented for not only degree and community structure, but also centrality structures. Indeed, generated graphs are only as accurate as their underlying models, and improvements in comparisons of graph topology may be possible with additional augmentation.

This was evidenced by our slightly biased estimate of the odds ratio of a physician being in internal medicine or classified as a hospitalist based on their proportion of same speciality neighbors. Of note, while estimates of the odds ratio calculated on the generated graph were consistently over-estimated, they consistently fell within the range of the 95\% confidence interval of the estimate calculated on the original graph. The bias in the estimator is likely due to a loss of information between the original graph and generated graphs, where medical specialty was the only directly measured attribute used to generate the graph, which may have over-emphasized the importance of connections based on medical specialty (e.g., if medical speciality is correlated with other measures that are also partially responsible for the generation of the graph). Researchers generating complex graphs should take care to confirm that the combination of provided labels adequately represents the target relationships of interest. 

As well, the current framework weights all attribute and structural augmentation labels equally. Future directions will include optimizing a weighting function of possible labels using a regression framework in order to best capture information related to both network structure and vertex attribute relationships. 

\section*{Conclusions}

Graphs generated from real data have many possible use cases. Scalable graphs can be important for bench-marking tools and graphs generated from confidential data can be safely distributed with software and publications. Moreover, the distribution of generated biomedical graphs may reduce the costs of data acquisition for new researchers, allow greater access for trainees, reduce the risk of analyzing data with novel tools that violate attempts to analyze the data securely, provide possible datasets for validation studies, support reproducible research, and be useful in pilot analyses for hypothesis generation. GRANDPA is a flexible and user-friendly framework which will allow researchers across disciplines to generate meaningful graphs from real data.

\section*{Acknowledgements}
The authors thank Megan Murphy for extracting and providing access to the Medicare data used to illustrate the application of GRANDPA to a network whose edges contain partially confidential information, and the P01 team (grant number in Funding section) leadership for supporting this work through access to the Medicare data. The authors also thank Christian Darabos and Research Computing at ITC at Dartmouth College for advise, support, and computational infrastructure supporting these analyses. 


\section*{Funding}
This work was supported by the National Institutes of Health in the USA (grant number P01 AG019783). JL is supported by NIH grant subawards under R24GM141194, P20GM104416 and P20GM130454.

\section*{Abbreviations}
ATG: Attribute graph model \\
CCDF: Complimentary cumulative distribution function \\
CMS: Centers for Medicare and Medicaid Services. \\
ERGM: Exponential random graph model \\
MAG: Multiplicative attribute graph \\
GRANDPA: Generative networks using degree and property augmentation \\
NRMSE: Normalized root mean square error \\
PGM: Property graph models

\section*{Availability of data and materials}
Original data are available on request due to privacy or other restrictions. Simulated data is available at \url{https://github.com/CarlyBobak/grandpa}.

\section*{Competing interests}
The authors declare that they have no competing interests.

\section*{Authors' contributions}
CAB and AJO conceptualized and formulated the algorithm, and identified the Medicare Unipartite Physician Network case-study. CAB and YZ identified the Zachary Karate Club case study, created the R package, finalized the figures, and drafted the initial manuscript. JJL evaluated the code and documentation. JJL and AJO provided guidance and insights during manuscript development and editing. All authors read and approved the final manuscript.

\bibliographystyle{unsrt}  
\bibliography{references}  


\end{document}